# MUNDOS VIRTUALES COMO APOYO A LA DOCENCIA EN INGENIERÍA

# VIRTUAL WORLDS AS A SUPPORT TO ENGINEERING TEACHING

**Roberto Muñoz**[1,2]   **Marta Barría**[1]   **Cristian Rusu**[2]

**RESUMEN**

Los Mundos Virtuales (MVs), son una tecnología emergente utilizada cada vez más por un mayor número de instituciones educativas de todo el mundo. Se trata de un entorno, un medio de aprendizaje y una herramienta educacional que permite distintos niveles de interacción en línea. En la cátedra de "Programación I", de la carrera de Ingeniería Civil en Informática de la Universidad de Valparaíso, se realizó una experiencia piloto utilizando el MV de *Second Life*, con el fin de evaluar el potencial del uso de MVs en la práctica docente.

*Palabras clave: Mundos Virtuales, Second Life, Educación Virtual, Educación a Distancia, Enseñanza en Línea.*

**ABSTRACT**

Virtual Worlds (VWs) are an emerging technology used by a growing number of educational institutions around the world. It is an environment, a way of learning and an educational tool that allows different levels of online interaction. In the course "Programming I", of the career Informatics Engineering at Universidad de Valparaíso, we conducted a pilot experience with the VW of *Second Life*, in order to evaluate the potential of using VWs in the teaching practice.

## INTRODUCCIÓN

La innovación en la enseñanza y el aprendizaje en la educación superior, está siendo impulsada por la creciente disponibilidad y potencialidad de las Tecnologías de Información y Comunicación (TIC). Las TIC ofrecen la posibilidad de que nuevas formas de enseñanza sean implementadas más allá de la sala de clases tradicional [1].

Una de las nuevas formas de enseñanza, viene dada a través de los Mundos Virtuales (MVs). Éstos no son algo desconocido, puesto que existían mucho antes de la expectación que se creó en torno a *Second Life* (SL) [2].

Hoy en día, los MVs son espacios donde cientos de personas pueden interactuar con otros, de manera simultánea en un entorno simulado computacionalmente, en dos o tres dimensiones [3].

Cada MV requiere que los usuarios elijan o creen avatares, los que le permitan interactuar con objetos, con el entorno virtual y con otros usuarios (a través de sus avatares) [4]. La identidad del avatar (edad, género, y raza), puede diferir de la identidad del usuario. El principal medio de comunicación es el texto, pero la voz, los gestos y las expresiones están también disponibles [5].

Los MVs 3D, como SL, ofrecen plataformas con distintos niveles de interacción. Mientras que antes de la introducción de los MVs el aprendizaje en línea se limitaba sólo a las formas textuales sincrónicas y asincrónicas, tales como foros, chat, y cuestionario-respuesta, los MVs ofrecen un mayor nivel de participación, cara a cara, en el proceso de enseñanza-aprendizaje, tal como ocurre en las salas de clases tradicionales.

Estos entornos, despertaron el interés de educadores e instituciones de todo el mundo, con más de 750 Universidades ya ubicadas en la isla de SL [6]. Los MVs, se han consolidado como espacios valiosos de aprendizaje, en el que las casas de estudios están explorando otras formas de inmersión en educación, a través de actividades que se encuentran en el límite de lo real y virtual, incluyendo muchas veces simulaciones con realidad aumentada [7].

Los MVs, no son una panacea para la educación superior, pues presenta muchos desafíos para los estudiantes y docentes. Este artículo describe cómo SL se ha utilizado para la enseñanza y el aprendizaje en la cátedra de Programación I de la carrera de Ingeniería Civil en Informática de la Universidad de Valparaíso. Se analizan las oportunidades, retos y problemas en el uso de los MVs. Además, de examinar los datos recogidos de una encuesta realizada a los estudiantes que utilizaron como sala de clases virtual en SL para su aprendizaje.

---
[1] Escuela de Ingeniería Civil en Informática, Facultad de Ingeniería. Universidad de Valparaíso. Código Postal 2360102. Valparaíso, Chile. {roberto.munoz.s, marta.barria}@uv.cl.
[2] Escuela de Ingeniería Informática, Facultad de Ingeniería. Pontificia Universidad Católica de Valparaíso. Código Postal 2362807. Valparaíso, Chile. roberto.munoz.s@mail.pucv.cl, cristian.rusu@ucv.cl.

El siguiente documento se estructura de la siguiente forma: En la Sección I, son presentados los Tipos de MVs existentes, junto con sus principales características. En la Sección II, se contextualiza a la enseñanza en MVs. En la Sección III, se describe a la Plataforma utilizada en esta experiencia. En la Sección IV, se presenta la Metodología empleada, para luego en la Sección V, detallar el Caso de Estudio. En la Sección VI, se presenta los Resultados del experimento. Para finalizar con las Conclusiones y Trabajo Futuro de esta investigación.

## TIPOS DE MUNDOS VIRTUALES

Los MVs, en la actualidad, poseen una amplia gama de aplicaciones, tales como: educación, entretenimiento, comercio electrónico, investigación científica, entre otros. No existe una única y ampliamente aceptada clasificación de MVs. Sobre la base propuesta por [8], los autores de [9], plantearon un conjunto de criterios a fin de establecer la tipología de MVs. Éstos fueron refinados por [10], para aplicarlos en educación. Los MVs se pueden clasificar de acuerdo a los siguientes criterios:

- *Propósito* (contenido de la interacción): Consiste en determinar cuál es el foco del MV (temático, específico a una comunidad, para niños, educativos, de objetivo abierto etc.). Aplicado a la educación en MV, este criterio corresponde a si el MV está orientado para dar soporte, por ejemplo a: seminarios virtuales, aprendizaje experimental, resolución de problemas, diseño, construcción, entre otros.
- *Lugar* (localización de la interacción): El lugar de interacción puede ser completamente o parcialmente virtual, considerando también la distribución geográfica de los usuarios. Aplicado a la educación en MVs, esto corresponde a indicar si la interacción será realizada en su totalidad en el MV o bajo un enfoque de aprendizaje mixto.
- *Plataforma* (diseño de la interacción): En los MVs asociados a educación, la elección de la plataforma, determinará la naturaleza de las posibles interacciones que ocurrirán como su flexibilidad para el trabajo colaborativo.
- *Población* (modelo de la interacción): Corresponde al tamaño del grupo, sus características sociales, como también las características del mercado objetivo del MV. En educación, esto se podría distinguir en el nivel educativo de los participantes (básica, media, superior, postgrado), o bien en el campo disciplinario que ocurre la interacción (Ingeniería, Medicina, Leyes, u otra).
- *Modelo de Negocios* (retorno de la interacción): Corresponde al modelo de ganancia del MV. En educación, esto puede ser asociado a si se cancela un arancel o no, relacionado al curso o carrera impartida en el MV.

Adicionalmente, se ha podido detectar que todos los MVs comparten las siguientes características [11]:

- *Avatar*: Cada usuario es representado por su (único) avatar, el cual es una representación digital del usuario en el MV. Éstos son parte central de cualquier mundo virtual [12].
- *Reglas del Mundo Virtual*: Un MV posee sus propias reglas (físicas) inviolables.
- *Espacio Compartido*: Un MV permite a numerosos usuarios participar en el mismo de manera simultánea.
- *Interacción y Comunicación:* Se realizan en tiempo real entre usuarios (por medio de sus avatares).
- *Persistencia:* Un MV es (parcialmente) persistente, independientemente si los usuarios se encuentran conectados o no.
- *Personalización:* Los MVs permiten a los usuarios modificar, desarrollar, construir, o generar contenido personalizado.
- *Entorno Gráfico:* Un MV es representado computacionalmente de manera gráfica, en entornos que van desde 2D a 3D.

## ENSEÑANZA EN MUNDOS VIRTUALES

El aprendizaje es un fenómeno social, donde los estudiantes adquieren los elementos necesarios para apropiarse del conocimiento a través de la interacción con los pares, profesores y el material.

Para la educación con TIC, es necesario recurrir a los principios teóricos generales y percepciones acerca la enseñanza y el aprendizaje, pero ahondar más en cómo realizarlas. Desde una perspectiva socio-cultural, el aprendizaje es visto como un proceso mediante el cual se produce una interacción social entre dos o más personas que cooperan en una actividad conjunta, con el propósito de producir un conocimiento [13].

La enseñanza y aprendizaje, se produce más fácilmente en situaciones colectivas y colaborativas. Dentro de las salas de clases tradicionales, esto puede ser facilitado mediante trabajos en grupo, juegos, entre otros. El aprendizaje en los MVs es mucho más que un fenómeno social, puesto que favorecen el aprendizaje colaborativo y la construcción de conocimiento, a través de la interacción sociocultural entre los involucrados [14].

Desde su nacimiento, las plataformas de MVs se concibieron como puntos de interacción online entre personas a través de personajes (avatares), que se desenvuelven en espacios en dos o tres dimensiones. En la actualidad, estos mundos están sufriendo una reinvención y especialización hacia la formación online.

Los MVs ofrecen las ventajas de la educación a distancia, en un entorno tridimensional en la que es posible interactuar tanto con docentes, como con alumnos. Lo que propicia el aumento de la participación y la implicación en el proceso educativo.

## PLATAFORMA

SL, es un entorno virtual en tres dimensiones liberado el 23 de junio de 2003, desarrollado por Linden Lab, y es accesible gratuitamente en Internet [15]. Sus usuarios, conocidos como "residentes", pueden acceder a SL mediante el uso de uno de los múltiples programas llamados *Viewers* (visores), lo cual les permite interactuar entre ellos mediante un avatar. Los residentes pueden así explorar el mundo virtual, interactuar con otros residentes, establecer relaciones sociales, participar en diversas actividades tanto individuales como en grupo, crear y comerciar propiedad virtual y servicios entre ellos. SL está reservado para mayores de 16 años [16].

Este MV, se compone de regiones, que son tierras independientes de 256x256 metros. Cada región puede tener un máximo de cuatro regiones adyacentes, las que pueden ser públicas o privadas. Las regiones públicas son propiedad de Linden Lab, mientras que las regiones privadas, son adquiridas por particulares o empresas [17].

La apariencia de una región se define por los objetos que la componen. Cada región tiene una política específica sobre la creación y destrucción de objetos. Por ejemplo, las regiones *Sandbox* (Cajas de arena), son lugares utilizados por los avatares para probar sus nuevos objetos, los que son destruidos después de su creación.

SL, permite una comunicación fluida, colaboración virtual y creación de contenido 3D. Este MV, ofrece un gran potencial para la enseñanza y el aprendizaje.

Existen muchas posibilidades para la aplicación de SL en un contexto educativo formal. Entre ellas están la reproducción de una facultad en un MV, en la que se fomente el sentido social de los estudiantes, o bien en la que se anime a los estudiantes a exponer sus experiencias. También puede utilizarse como punto de reunión, en la que la distancia no sea un impedimento.

Cabe mencionar que las actividades que se desarrollan en SL tienen como eje central la interacción entre personas de todo el mundo, y cada vez más las acciones formativas, sociales y culturales, están encontrando en SL un entorno propicio para cambiar las formas de acceso a la información y comunicación [18].

La elección de este MV, se debió a que ofrece una nueva dimensión en lo que respecta interacciones en línea, puesto que los usuarios no sólo se comunican a través de texto en una pantalla, abstraídos totalmente de de la realidad, sino que también interactúan a través de representaciones visuales de los mismos, por medio texto y audio en una realidad alternativa [19].

## METODOLOGÍA A UTILIZAR

La metodología utilizada para este experimento contó con los siguientes pasos.

- Realización de conferencia apoyada por recursos audiovisuales y ejemplos prácticos, acerca de qué son los MVs, principales componentes, características y tipos. Esta conferencia fue realizada de manera presencial a los alumnos interesados en participar en esta experiencia.
- Capacitación en la utilización de este tipo de entornos, en particular con la plataforma SL. Esta capacitación se realizó en el laboratorio de Gestión de la Escuela de Ingeniería Civil en Informática de la Universidad de Valparaíso [20], en la que los alumnos interactuaron con este MV. Se realizó de manera grupal, con el propósito de resolver de manera colaborativa, consultas y/o dudas con respecto a: su utilización, trabajo con SL, creación de avatares por parte de los alumnos, entro otros.
- Visita de los alumnos por medio de sus avatares al campus virtual de la Universidad de Sevilla (US) [21].
- Capacitación en la utilización de estos elementos y otros virtuales para poder utilizar en una clase de Programación en el lenguaje C (ej. Utilización de un cuaderno de notas virtual para poder enviar los ejercicios realizados en clases).
- Entrega de elementos básicos de interacción, entre los que destaca un set de gestos y expresiones específicas (ej. gestos de afirmación, negación, expresiones de aburrimiento, consulta, entre otros). Estos elementos fueron desarrollados por la US, para su utilización en una sala de clases. El set de gestos básicos se puede apreciar en la Figura 1.

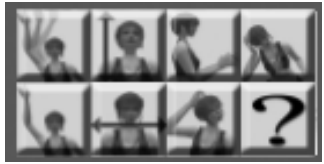

Figura 1. Set de gestos básicos US en SL.

- Realización de tres clases virtuales teórico – prácticas de 1,5 horas cada una.
- Realización de una encuesta que contó con preguntas abiertas y cerradas.
- Evaluación de la Unidad vista en clases virtuales por medio de un Laboratorio práctico.
- Análisis y publicación de los resultados.

Esta metodología fue aplicada con el caso de estudio presentado en el punto posterior.

## CASO DE ESTUDIO

Para llevar a cabo la experiencia, se contó con 20 alumnos de la cátedra de Programación I, de la carrera de Ingeniería Civil en Informática de la Universidad de Valparaíso. Los alumnos debieron asistir en forma virtual al campus en SL que facilitó la US [21]. En la Figura 2, se observa el auditorio facilitado para realizar la experiencia.

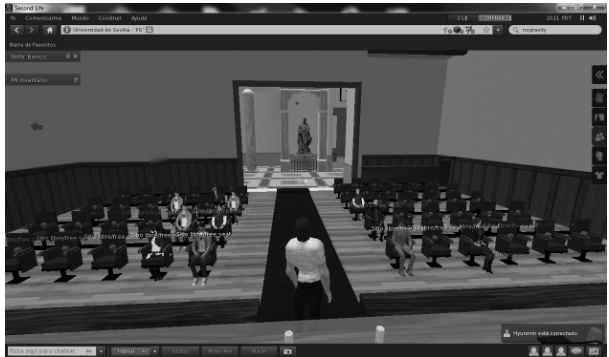

Figura 2. Auditorio Universidad de Sevilla en SL.

La Unidad a tratar fue Estructuras en el Lenguaje C, la que fue vista en tres clases virtuales de 1,5 hrs cada una. Ésta fue apoyada por recursos visuales (presentaciones virtuales), además de la realización de ejemplos y ejercicios prácticos.

Posterior a la realización de estas 3 clases, se efectuó una encuesta de 10 preguntas, la que tenía como objetivo analizar el aporte de SL a la educación y formación académica, junto con capturar las impresiones de los alumnos frente a este nuevo medio de aprendizaje. Además, se realizó un laboratorio práctico de 1,5 hrs. para evaluar la Unidad vista en las clases virtuales.

## RESULTADOS

Se desarrolló un laboratorio práctico de estructuras en C, en el que se evaluó puntos tales como: conceptos básicos, funciones, arreglos y ordenamiento de estructuras.

Se asume que tanto presencial como virtualmente hay factores externos que influyen en el rendimiento de los alumnos de primer año, tales como motivaciones, horas dedicada al estudio, nivel de asistencia, entre otros [22]. Las calificaciones obtenidas en años anteriores en esta misma unidad de la cátedra, dictadas por el mismo docente y bajo la misma metodología de enseñanza fue contratada con aquellas obtenidas bajo el entorno SL (ver Tabla 1), observándose lo siguiente:

- Usando SL el promedio del curso, aumentó en aproximadamente un 11%, comparándolo con el promedio obtenido en los 2 años anteriores (año 2009 y 2010 de la Tabla 1)
- La distribución de las calificaciones, tuvo una desviación estándar aproximadamente igual a la de años anteriores.
- La máxima calificación obtenida se mantuvo con respecto a los años anteriores y la mínima mejoró.

Tabla 1. Histórico calificaciones Programación I.

|            | 2009 | 2010 | 2011 (MV) |
|------------|------|------|-----------|
| Promedio   | 4.7  | 4.1  | 4.9       |
| Max.       | 7.0  | 7.0  | 7.0       |
| Min.       | 1.0  | 1.0  | 2.0       |
| Desv. Est. | 2.2  | 2.1  | 2.2       |

Con respecto a los resultados de la encuesta, se puede mencionar lo siguiente:

- El 80% de los encuestados, participó activamente en todas las clases o aportaba cuando tenía consultas. El 20% restante, prefería escuchar.
- El 85% no percibió diferencias entre asistir a una clase virtual y a una presencial. El 15% restante, advirtió leves diferencias.
- El 100% de los encuestados, consideró que el uso de MVs es un apoyo a su formación académica.
- El 100% de los encuestados, repetiría la experiencia con otra unidad del ramo.

## CONCLUSIONES Y TRABAJO FUTURO

Las posibilidades educativas y los cambios en los procesos que se puede generar en este tipo de entornos soy muy importantes, puesto que las distancias y costos no son un impedimento en los MVs.

Los resultados favorables del experimento, se pueden deber a que quienes asistieron a estas cátedras virtuales fueron voluntarios, pero no por ello, los resultados debieran ser menos representativos.

Los resultados muestran que los alumnos que aprendieron usando el MV, obtuvieron notas en promedio 11% mejores que las obtenidas por los alumnos que no utilizaron este medio. Sin embargo el objetivo de este estudio fue demostrar la factibilidad de la realización de este tipo de actividades en entornos virtuales como medio de apoyo a la enseñanza presencial.

El entorno virtual de SL, ofrece una oportunidad única para que los educadores lleven a cabo, en su plan de estudios, actividades interactivas en tiempo real que apoyen el aprendizaje colectivo.

Al incorporar estas herramientas, entran en juego diferentes dimensiones como: el de la didáctica, el de la disciplina que se pretende enseñar y el de la disciplina que transportan.

Se espera realizar más experiencias de este tipo en cátedras introductorias, tales como Fundamentos de Programación, y analizar el comportamiento académico bajo este tipo de entorno virtual como medio de apoyo.

Si lo que se desea es emplear SL para una clase en formato tradicional (profesor explicando y alumno escuchando), no es el lugar más apropiado, SL obliga al docente replantear nuevas formas de enseñanza y aprendizaje utilizando TICs.

Las posibilidades educativas y los cambios en procesos que puede llegar a generarse, bajo este entorno, son muy importantes. No se debe dejar que este momento ocurra sin ser partícipes de ello, siendo artífices, creadores y promotores.